\documentclass[useAMS,usenatbib]{mn2e}
\usepackage{graphicx}
\usepackage{epstopdf}
\usepackage{bm}
\usepackage{times}
\usepackage{amsmath}
\usepackage{color}
\usepackage{hyperref} 
\newcommand{\adsurl}[1]{\href{#1}{ADS}} 
\providecommand{\url}[1]{\href{#1}{#1}}

\begin{document}

\title[Reply to two Comments on ``The Contribution of potassium (and Chlorine) to the 3.5 keV line'']{Reply to Two Comments on ``Dark matter searches going bananas the contribution of Potassium (and Chlorine) to the 3.5 keV line''}

\author[T. Jeltema and S. Profumo]{Tesla Jeltema$^{1}$\thanks{tesla@ucsc.edu} and Stefano Profumo$^{1}$\thanks{profumo@ucsc.edu}\\
$^{1}$Department of Physics and Santa Cruz Institute for Particle Physics
University of California, Santa Cruz, CA 95064, USA
}

\maketitle

\begin{abstract}
We respond to two comments on our recent paper, \citet{bananas}. The first comment by Boyarsky et al. confirms the absence of a line from M31 in the 3-4 keV energy range, but criticizes the choice of that energy range for spectral fitting on the basis that (i) the background model adopted between 3-4 keV is invalid outside that range and that (ii) extending the energy range multiple features appear, including a 3.5 keV line. Point (i) is manifestly irrelevant (the 3-4 keV background model was not meant to extend outside that range), while closer inspection of point (ii) shows that the detected features are inconsistent and likely unphysical. We demonstrate that the existence of an excess near 3.5 keV in the M31 data requires fitting a broad enough energy range such that the background modeling near 3.5 keV is poor to a level that multiple spurious residual features become significant.
Bulbul et al. criticize our use of WebGuide instead of the full AtomDB package. While a technically correct remark, this is only a red herring: our predictions are based on line ratios, and not on absolute emissivities; line ratios, for atomic transitions with similar peak temperatures, are largely temperature-independent, therefore the line ratios we employed to draw our conclusions are substantially correct. Bulbul et al. also present a new analysis of their data at lower energy, which excludes a significant Cl contamination to the 3.5 keV line. This is a welcome new element, but irrelevant to our main criticism of their analysis, since Cl XVII emission was predicted to be subdominant to K XVIII. Both of the Bulbul et al.'s criticisms are thus entirely inconsequential to the conclusions we drew in our original study. Finally, we demonstrate that the multi-temperature models employed in Bulbul et al. are, in fact, inconsistent, based on the Ca XX to Ca XIX line ratio, which is a solid test,  independent of relative elemental abundances; we show that the overestimated cluster plasma temperatures they employ lead to gross underestimates of the K XVIII line emissivity.
\end{abstract}

\begin{keywords}
dark matter -- line: identification -- Galaxy: centre -- X-rays: galaxies -- X-rays: galaxies: clusters
\end{keywords}   

\section{Introduction}

In 2009, one of the co-authors of \citet{Bulbul:2014sua} reported the discovery of an unidentified 2.5 keV line from the supposed ultra-faint dwarf spheroidal galaxy Willman 1 with {\em Chandra} \citep{Loewenstein:2009cm}. Excitement and model building efforts ensued, motivated by the possibility that this line originated from the radiative decay of 5 keV sterile neutrinos into an active neutrino and a photon; the fact that the resulting sterile neutrino particle properties would potentially also explain pulsar kicks was, in the words of the authors, additionally ``{\em bolstering both the statistical and physical significance of [the] measurement}''. Boyarsky and collaborators subsequently questioned the sterile neutrino interpretation of the 2.5 keV signal, based, notably, on the {\em non-detection of any lines from M31}, and outlined a {\em vade mecum} on how to ``{\em check the dark matter origin of a spectral feature}'' \citep{Boyarsky:2010ci}. \citet{Kusenko:2010sq} then promptly proceeded to criticize Boyarsky et al's findings, principally based on the dark matter density profile choice for M31. Eventually, new XMM observations \citep{Loewenstein:2012px} conclusively showed that no statistically significant line existed from Willman 1.

Earlier this year, \citet{Bulbul:2014sua} reported the detection of an unidentified X-ray line at an energy of about 3.5 keV from various  observations and different sub-samples of stacked clusters of galaxies, and entertained the possibility that the unidentified line originated from sterile neutrino decays, short of any plausible, more mundane astrophysical explanation. Shortly thereafter, \citet{Boyarsky:2014jta} confirmed the existence of a 3.5 keV line from the Perseus cluster and, interestingly, from M31. Concurrently with significant model building work, other investigators  scrutinized X-ray data to further seek confirmation for, and/or to constrain, the 3.5 keV line. \citet{Riemer-Sorensen:2014yda}, using Chandra observations of the Galactic center, did not find any evidence for a line at 3.5 keV in excess of atomic emission from K XVIII, setting constraints that started to cast doubts and concerns about the possible sterile neutrino origin of the line. 

As a famous quote from George Bernard Shaw kept lurking in the back of our minds (``{\em We learn from history that we learn nothing from history\footnote{Shaw actually attributes this quote to G.W.F. Hegel.}}''), in \citet{bananas} we embarked in the following three tasks: (i) exploring deeper observations of the Galactic center with XMM, (ii) re-examining the archival M31 observations analyzed in \cite{Boyarsky:2014jta}, and (iii) critically probing the claims that no astrophysical explanation existed for the 3.5 keV line observed from clusters of galaxies. The gist of our findings is as follows:
\begin{itemize}
\item[(i)] We discovered an X-ray line at 3.5 keV in the GC; the line flux is compatible with a sterile neutrino decay interpretation of the 3.5 keV line from other systems\footnote{This point seems to have been at least partly overlooked or under-stated in the way our work was subsequently quoted in the literature.}; however, the 3.5 keV line flux was found to also be compatible with the expected emission from two atomic transitions of K XVIII within systematics.\\
\item[(ii)] We did not find any statistically significant line at 3.5 keV in the X-ray spectrum from M31 in the energy range between 3 and 4 keV.\\
\item[(iii)] We argued that the algorithm adopted in \citet{Bulbul:2014sua} to estimate the K XVIII emission was possibly not conservative enough, focussing on the issue of them employing temperature models systematically biased towards large temperatures; we also pointed out that \citet{Bulbul:2014sua} did not include, without any justification, a possible additional atomic transition line from Cl XVII; we advocated for a more conservative and generous approach towards estimating the potential background from K XVIII (and possibly from Cl XVII) as an explanation for the 3.5 keV signal, and claimed that within a more conservative estimate for the brightness of elemental lines, the observed signal strength might be compatible with, exclusively, the flux from known atomic transition lines.
\end{itemize}

After our study had appeared, three additional analyses tested the sterile neutrino decay interpretation of the 3.5 keV line with observations of different astrophysical objects. In \cite{Malyshev:2014xqa} archival XMM observations of selected local dwarf spheroidal galaxies were employed to exclude with robust statistical significance a dark matter decay origin for the 3.5 keV line reported by \citet{Bulbul:2014sua}. Again, Shaw's quote comes to mind as one returns to a paper by one of the co-authors of \citet{Bulbul:2014sua} replying to the lack of a 2.5 keV line signal from M31 reported by \citet{Boyarsky:2010ci}, which stated that: ``{\em M31 is not as good a target for future observations as dwarf spheroidal galaxies. The ongoing and planned observations of dwarf spheroidal galaxies give the best opportunity to confirm or rule out 5 keV sterile neutrinos as a dark matter candidate.}" \citep{Kusenko:2010sq} Intriguingly, such observations of dwarf spheroidal galaxies, now archival, rule out with robust statistical significance \citep{Malyshev:2014xqa} the sterile neutrino parameter space favored by a dark matter decay interpretation of the results in \citet{Bulbul:2014sua}.

Another study that cast serious doubt on a dark matter decay interpretation of the 3.5 keV line, \citet{Anderson:2014tza}, analyzed stacked observations of a large number of galaxies and galaxy groups, where little plasma emission is expected above 2 keV, employing both Chandra and XMM observations. Again, no signal was detected at 3.5 keV, ruling out the dark matter decay interpretation of the results of \citet{Bulbul:2014sua}
at the 4.4$\sigma$ level using Chandra data and at the 11.8$\sigma$ level with the XMM stacked observations.

A third recent study \citep{Urban:2014yda} analyzed deep Suzaku observations of Perseus and of three additional X-ray bright galaxy clusters, Coma, Virgo and Ophiuchus. \cite{Urban:2014yda} finds evidence for a 3.5 keV line in Perseus, but with a radial profile inconsistent with expectations from dark matter decay. More crucially, \cite{Urban:2014yda} {\em does not find} any evidence for a 3.5 keV signal from Coma, Virgo or Ophiuchus commensurate with the Perseus signal. Also, \cite{Urban:2014yda} confirms our original claim that it is problematic to find a single temperature to the plasma based on line strengths (unlike what argued in \cite{Bulbul:2014ala}), and shows that the 3.5 keV line feature in Perseus may in fact have an elemental origin \citep[as we also claimed in][]{bananas}.

Finally, we recently showed, in collaboration with Eric Carlson, that the morphology of the 3.5 keV emission from the Galactic center and from the Perseus cluster is incompatible with a dark matter decay or annihilation origin, or with axion-like particle conversion \citep{carlson}: the radial and azimuthal signal distribution is distinctly at odds with what expected from dark matter decay or conversion; the Galactic center emission is correlated with emission from bright atomic transition lines, and clearly lacks azimuthal symmetry; the emission from Perseus is associated with the cluster's cool core and drops sharply in the cluster's outskirts \citep[in complete agreement with what found in][with Suzaku data]{Urban:2014yda}. Finally, our binned likelihood analysis allowed us to derive the most stringent constraints to-date on a dark matter decay interpretation of the 3.5 keV signal which excludes with high statistical confidence a dark matter decay origin for the signal reported in \citet{Bulbul:2014sua} and \citet{Boyarsky:2014paa}.

In this note, we address the concerns expressed in the two ``comments''  on our original study \citep{bananas} that have appeared recently, namely \citet{Boyarsky:2014paa} (which we address in the following section \ref{boyarsky}) and \citet{Bulbul:2014ala} (which we address in section \ref{bulbul}). We demonstrate here that neither comment affects any of the key findings (i), (ii) and (iii) listed above,  and we point out a few additional issues and inconsistencies in the algorithm employed to estimate the flux of the K XVIII lines in \citet{Bulbul:2014sua} and in the findings of \citet{Boyarsky:2014paa}, casting additional concern on the validity of those results.

\section{Summary and reply to the Comment by Boyarsky et al}\label{boyarsky}

In short, the reason why the 3.5 keV line appears in a broad (2-8 keV) energy window but not in the 3-4 keV window is that as one extends the energy range, the continuum background gets increasingly complex to the point that a simple power-law is no longer sufficient to characterize it; additional spurious features then appear even with a more sophisticated background model, as an obvious result of the poor background characterization. We show below in detail that the arguments produced by \citet{Boyarsky:2014paa} to disqualify the choice of employing the 3-4 keV window are flawed; that even employing a broader energy range no statistically significant 3.5 keV line is present; and, finally, that the line features  \citet{Boyarsky:2014paa} claim to detect in the 3-4 keV range are inconsistent with atomic transition processes and very likely unphysical.

The key criticisms raised by \citet{Boyarsky:2014paa} are as follows:

\begin{enumerate}
\item[(1)] Restricting the energy range for the M31 analysis to the 3-4 keV is not justified because extending to higher energies the background model that fits the 3-4 keV range over-predicts count rates above 4 keV;
\item[(2)] The significance of the detected line is greater using a wider energy window because the constraints on the background parameters are tighter when one employs a broader energy range; limiting the analysis to the 3-4 keV interval results in increased uncertainty and in artificially decreasing the significance of the detection;
\item[(3)] Additional known atomic line complexes can be identified between 3 and 4 keV. The fluxes of the additional atomic transition lines strongly disfavor a K XVIII origin for the 3.5 keV line;
\item[(4)] A consistent interpretation of the 3.53 keV line as emerging from K XVIII in all objects is problematic.
\end{enumerate}

To start with, \citet{Boyarsky:2014paa} reproduces our findings when analyzing the M31 spectrum in the 3-4 keV range and the fact that there is no statistically significant line at 3.5 keV. 

In the first of their key remarks, \citet{Boyarsky:2014paa} show that the best-fit power-law in the 3-4 keV range over predicts counts above 4 keV. This is of course a completely moot point, as the background model in the 3-4 keV was not devised to fit count rates outside that energy range! In searching for a line near 3.5 keV, one needs to model well the continuum near 3.5 keV, but a 3.5 keV line does not care what the continuum is at 7 keV.  The claim that using a 3-4 keV range is unjustified based upon this argument is thus manifestly unfounded. 

\citet{Boyarsky:2014paa} then argue that considering a broader energy range, the power-law parameters are determined with better precision (point (2) above). They proceed to point out that the precision with which the power-law parameters can be fit in the 3-4 keV range (about 3-6\%) is comparable to the relative strength of the supposed 3.5 keV line (about 4\%); on the contrary, the power-law parameters for the 2-8 keV range are constrained to better than 1\%. As a result, the lines in the 3-4 keV were ``{\em partially compensated by the power-law continuum}''.

\begin{figure}
\begin{centering}
\includegraphics[width=1.0\columnwidth]{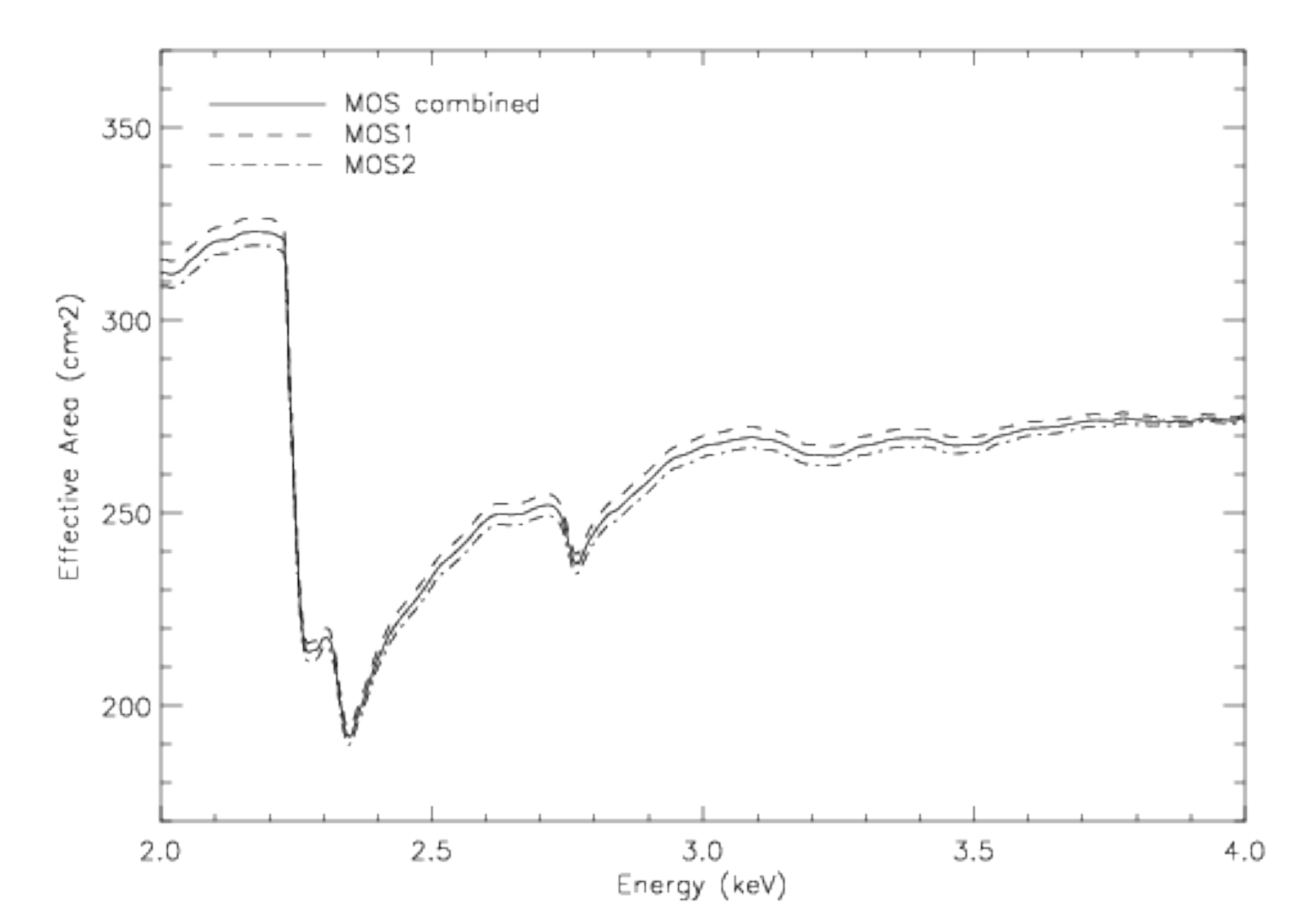}
	\end{centering}
\caption{Plot of the co-added ARF for M31 between 2 and 4 keV showing the variation in instrumental response across this range. Also shown are the co-added ARFs for the MOS1 and MOS2 detectors separately showing the slight difference in response between the two instruments.}
\label{fig:m31arf}
\end{figure}

First, a single power-law continuum background, even with arbitrary parameters, cannot ``compensate'' for a line, given that the 3-4 keV energy range is roughly one order of magnitude larger than the instrumental energy resolution. This can easily be tested by adjusting the power law parameters within their errors for the 3-4 keV fit: no excess 3.5 keV line appears.  Secondly, the better determination of the power-law parameters simply indicates that more data are being included and fit for, not that the quality of the fit to the background is better, which is the real physical question at stake. In fact, employing exclusively a power law fit to the 3-4 keV spectrum leads to a reduced chi-squared of 1.1, while using the significantly more complicated model of \citet{Boyarsky:2014jta} across the 2-8 range produces a reduced chi-squared of 1.3. 

The background structure for the 2-8 keV range is significantly more complex: for example, it includes several bright instrumental and astrophysical lines and a much larger particle background.  In fact, achieving a reasonable fit in this energy range  requires at a minimum a second unfolded power law to model the particle background, plus $\sim$ 8-13 lines giving at least 22 fit parameters instead of 2.  And, as discussed below, the 2-8 keV background model does a particularly poor job of modeling the data in the 2.8 to 3.8 keV region, producing spurious signals and unphysical conclusions.

\begin{figure*}
\begin{centering}
	\includegraphics[width=1.4\columnwidth]{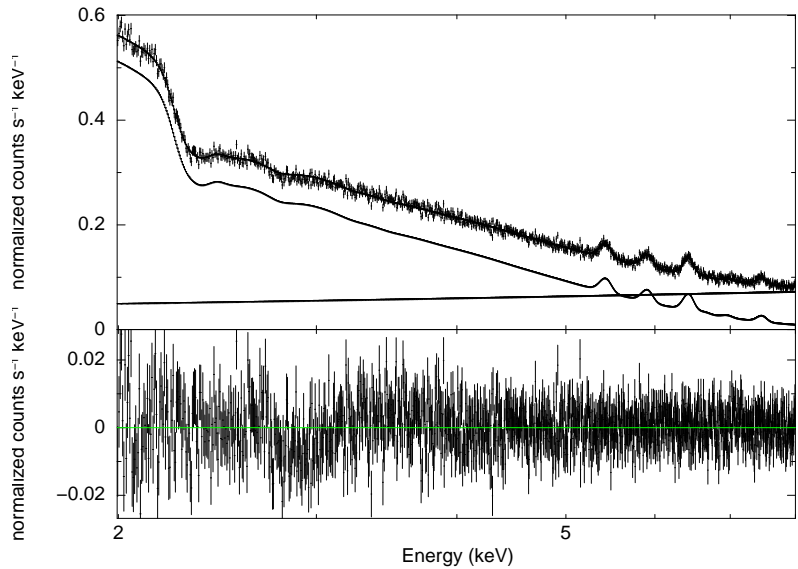}
	\end{centering}
\caption{{\bf Spurious residual features arise when modeling M31 over an unnecessarily broad energy range}. We show the best-fit model and residuals for the co-added spectrum of M31 between 2 and 8 keV.  The model includes a power law plus a second unfolded power law to model the continuum and several instrumental and astrophysical lines, similar to the model employed by \citet{Boyarsky:2014jta}.  No 3.5 keV line is included in the model.  Significant residuals can be seen over a large 1 keV band which are not line-like in nature.  In particular, the model overestimates the data from roughly 2.8 to 3.1 keV and slightly underestimates the data is a broad region from roughly 3.1 to 3.8 keV.}
\label{fig:m31spec}
\end{figure*}

While a power law continuum will not mimic a line feature, one should consider energy resolution when choosing a fit energy range, to ensure a sufficient number of independent bins.  Around 3 keV, the energy resolution of the XMM EPIC detectors is $\sigma \sim 42$ eV (FWHM $\sim 100$ eV), much smaller than the 1 keV wide band we initially employed.  We tested that if the size of the energy window is doubled to consider energies between 3-5 keV, a single power law still provides an acceptable fit (or a double power law with no lines) and a line near 3.5 keV is still not present at better than $2 \sigma$. Extending the data to a larger energy range from 3-7 keV, the background includes instrumental emission lines from Cr, Mn, and Fe-K as well as astrophysical plasma Fe line emission, and we minimally model the continuum as a power law plus an unfolded power law for the particle background.  Again, no significant excess near 3.5 keV is found at more than $2 \sigma$. This energy window is vastly larger than the energy resolution.  

Additional issues and unwarranted complications to the background model arise if one extends the data range to 2-8 keV range, as done in \citet{Boyarsky:2014jta}. In particular, in the 2-3 keV band the effective area of the instrument is more strongly varying (see Fig.~\ref{fig:m31arf}), including a couple of sharp instrumental edge features, and for the low temperature plasma in M31, several astrophysical plasma lines from S and Si appear in this energy range. A fit to the co-added MOS data for M31 in the 2-8 keV range with no 3.5 keV line is shown in Fig.~\ref{fig:m31spec} (see \citet{bananas} for details on the data selection and processing).  The model is chosen to be similar to the one employed by \citet{Boyarsky:2014jta}.  While the overall reduced $\chi^2$ is good (1.13, $\chi^2=1335$ for 1181 DOF), it can be seen from the residuals in Fig.~\ref{fig:m31spec} that {\em the continuum is not well modeled in the 2.8 to 3.8 keV range}.  In particular, the model overestimates the data from roughly 2.8 to 3.1 keV and slightly underestimates the data is a broad region from roughly 3.1 to 3.8 keV (see also Fig.~1 in \citet{Boyarsky:2014jta}).  These residual features stretch over a large $\sim 1$ keV band and are not line-like in nature.  Instead, the residuals point to inaccuracies in the continuum modeling and/or imperfect modeling of the spectral response.  Adding a line near 3.5 keV gives a a $\Delta \chi^2 = 12.5$ simply because this partially compensates for the broad residual excess, but it does not entirely account for the excess nor improve the residuals as a whole. In fact, we quite remarkably find that {\em adding an unphysical absorption line near 2.9 keV improves the fit by a much larger factor of $\Delta 
\chi^2 = 52$ and reduces the significance of an excess line near 3.5 keV to $2 \sigma$}.  

We conclude that if one wishes to find an excess near 3.5 keV in the M31 data, it is necessary to fit a broad energy range such that the background modeling near 3.5 keV is poor and a residual feature becomes significant.  Modeling only the relevant energies between 3-4 keV on either side of the proposed line energy, the continuum can be well modeled by a two parameter simple power law with no excess line emission.
 
Moving on to point (3), \citet{Boyarsky:2014paa} claim the detection of a 3.91 keV complex (Ca XIX and Ar XVII), and of three additional lines at 3.14 (Ar XVII/S XV) and at 3.37 keV (Ar XVIII/S XVI/Cl XVI). They argue that somehow in the 3-4 keV fit these lines were ``compensated'' by the power-law continuum -- it is entirely mysterious to us how a power-law, characterized by a slope and a normalization, could possibly compensate for the three new lines, plus the 3.5 keV line. 

We now address how realistic the new purported lines are. First, we note that one of the ``lines'' found, at 3.37$\pm0.03$ keV, is almost 2$\sigma$ away from the two brightest lines in the Ar XVIII/S XVI complex, at 3.323 and 3.318 keV (more than an order of magnitude brighter at peak than the S XVI line at 3.355 keV). 

By closer inspection, the line ratios resulting from the putative detections in \cite{Boyarsky:2014paa} reveal multiple inconsistencies hard to reconcile by invoking relative elemental abundances:

\begin{itemize}
\item the 3.14 keV complex is dominated by Ar XVII, and the 3.37 keV complex is dominated by Ar XVIII, so that ratio is to a good approximation abundance-independent. The flux associated with the 3.14 keV complex is found to be 0.6 times the flux of the 3.37 keV complex, which indicates (independent of abundances, since both complexes are dominated by Ar) a temperature larger than 5 keV, completely unrealistic for Andromeda.
\item In the large-temperature regime favored by the Ar ratio mentioned above, the 3.37 keV complex is always much brighter than the 3.91 keV complex, while the line fluxes indicate the 3.91 keV to be more luminous
\item At all temperatures, the 3.91 keV complex should be fainter than the 3.14 keV complex, while the flux ratio indicates that the 3.91 keV complex is about a factor 2 brighter than the 3.14 keV line. Since the 3.91 keV complex is dominated by Ca XIX, this would indicate a super-abundance of Ca over Ar, compared to solar, of up to more than an order of magnitude, depending on temperature.
\item If indeed Ca is overabundant over Ar by about an order of magnitude, the observed ratio of the 3.91 to 3.37 lines can never be brought into agreement, not even at temperatures larger than 10 keV.
\end{itemize}

It is thus clear that the putative newly detected lines are largely inconsistent, even allowing for generous uncertainties in the relative elemental abundances. As we showed above, inaccurate background modeling over an extended energy range can easily produce spurious features including additional lines, to compensate for poor modeling of the continuum emission. No robust conclusions or inferences based upon those lines can thus be made on predicting the line flux for the putative 3.5 keV line (point (4) above).

Finally, we notice that employing a broader energy range results in a stronger look-elsewhere effect, which \cite{Boyarsky:2014paa} and \cite{Boyarsky:2014jta} neglect:  there is in fact a relatively large likelihood that a $\sim3\sigma$ excess arises as a statistical fluctuation; for example, considering the number of bins between 3 and 4 keV in \cite{Boyarsky:2014jta}, a $\sim3\sigma$ fluctuation at any energy has a probability of occurring of more than 5\%. Taking the look-elsewhere effect into account, the significance of the detection in \citet{Boyarsky:2014jta} is well below 2$\sigma$.

\section{Summary and reply to the Comment by Bulbul et al}\label{bulbul}

\begin{figure*}\label{lineratio}
\begin{centering}
	\includegraphics[width=1.4\columnwidth]{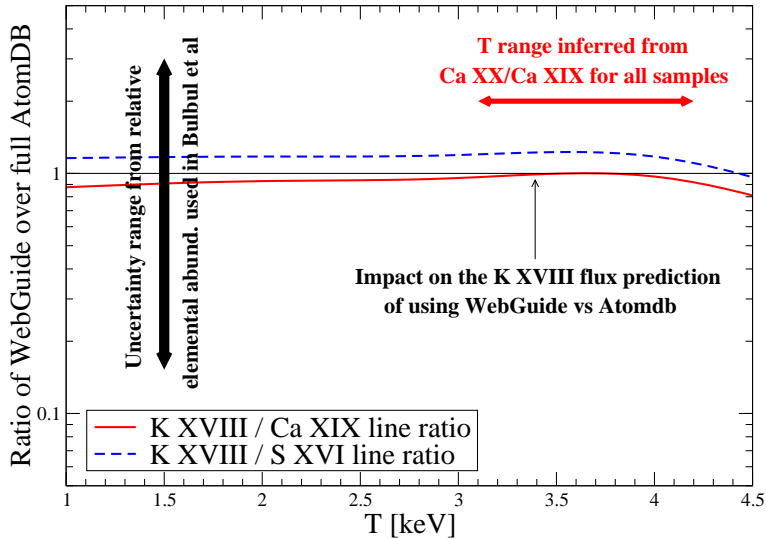}
	\end{centering}
\caption{{\bf Only a red herring: the impact of using the WebGUIDE tool versus the full AtomDB package is negligible compared to the size of systematic effects such as, for example, relative elemental abundances}. The red solid line and the blue dashed line illustrate the impact, as a function of temperature, and for temperatures relevant for our study, of using the approximate WebGUIDE emissivities versus the full AtomDB output in estimating the K XVIII line flux from reference lines. The lines correspond to the K XVIII to Ca XIX line ratio (red solid) and to the K XVIII to S XVI line ratio (blue dashed), and what is plotted is the WebGUIDE line ratio prediction divided by the full AtomDB prediction. We also show, for visual comparison, the range for the K XVIII line flux considered in the fits performed by \citet{Bulbul:2014ala}, and the temperature range for their cluster samples inferred from the (relative abundance-independent) Ca XX to Ca XIX ratio, across all samples and for both MOS and PN.} 
\label{fig:webguide}
\end{figure*}

In their comment to our work, \citet{Bulbul:2014ala} make three key points:

\begin{enumerate}
\item[(1)] In the calculation of the line ratios relevant to estimate the range for the K XVIII emission in \cite{bananas}, we employed the WebGUIDE tool instead of the full version of the AtomDB code; they proceed to argue that this would lead to ``very large errors'' in our estimates;
\item[(2)] They show that, in the specific case of the Perseus cluster, the temperature model used in their original analysis \citep{Bulbul:2014sua} produces a roughly consistent pattern of atomic emission lines (allowing for factors of order unity from relative elemental abundances), and insist that the upper limit to the K XVIII flux was estimated in a maximally conservative way;
\item[(3)] Finally, they extend the original spectral fit region to lower energies to include the 2.96 Cl XVII Ly$\alpha$ line; this line is a factor of at least $\sim6$ brighter than the Cl XVII Ly$\beta$ line at 3.51 keV. 
\end{enumerate}

Let us first examine point (1). \citet{Bulbul:2014ala} state in their comment that we claim to have used AtomDB while in fact they ``discover'' we had made use of the WebGUIDE tool. This is not accurate, as we include in footnote 2 the reference to the tool we actually used, WebGUIDE\footnote{http://www.atomdb.org/Webguide/webguide.php}. \citet{Bulbul:2014ala} correctly states that the predictions from the WebGUIDE tool differ from those obtained from the full AtomDB v2.0.2 database. However, this concern is quantitatively entirely irrelevant, as we show in figure \ref{fig:webguide}, and as evident from the bottom panel of their own figure 1 as well: the temperature dependence of the {\em line ratios}, which is what we employ for the predictions we quote in our Table 3, is negligible for all ratios with the exception of Ca XX (a line that peaks at significantly larger temperature than all other lines). The K XVIII prediction from the Ca XX line is however entirely irrelevant to our conclusions, as that line ratio does not significantly affect the range of possible K XVIII fluxes we base our conclusions upon. \citet{Bulbul:2014ala} is thus incorrect in stating that our analysis is ``{\em severely affected} by [the] use of the approximate atomic data''.

Using AtomDB in fact confirms what we had argued in \cite{bananas}, i.e. that for broad ranges of temperatures the predicted K XVIII flux from Ca XIX or Ca XX is compatible, within a factor of a few, with the observed flux associated with the ``unidentified'' 3.5 keV line for both the GC and clusters.  For example, comparing Table 1 in \citet{Bulbul:2014ala} with Table 3 in \cite{bananas} it can be seen that the predicted K XVIII line fluxes for the GC change by an insignificant amount except in the case of the Ca XX predictions (which, at low temperatures, significantly increase, thus making our original K XVIII flux predictions overly conservative).  Even ignoring the Ca XX predictions at low temperatures, the predicted K XVIII line fluxes are within a factor of 2 of the measured line flux over the full temperature range.

\begin{figure*}
\begin{centering}
	\includegraphics[width=1.8\columnwidth]{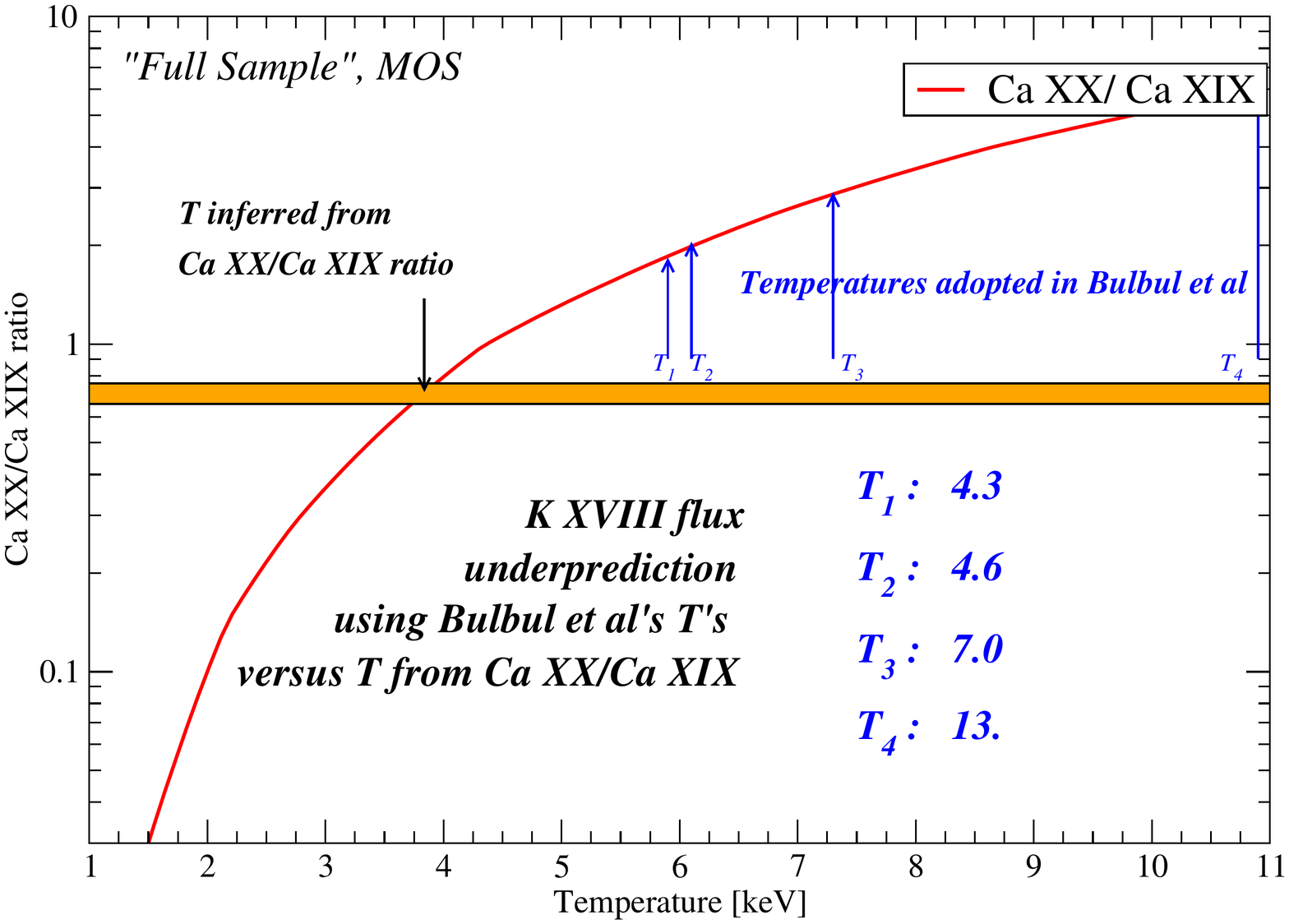}
	\end{centering}
\caption{{\bf Inconsistent multi-temperature models in {\protect\cite{Bulbul:2014sua}} lead to large under-estimates for the K XVIII line flux.} We show with a red line the theoretical prediction (from AtomDB v2.0.2) for the Ca XX to Ca XIX ratio as a function of temperature, while the line ratio as measured in {\protect\cite{Bulbul:2014sua}} for the full cluster sample MOS observations is shown in orange. The black vertical line indicates the resulting inferred plasma temperature. We indicate with blue arrows the four temperatures employed in the multi-temperature model utilized in {\protect\cite{Bulbul:2014sua}} to predict the K XVIII line flux. Using high temperatures leads to significant under-estimates of the K XVIII line flux, which is a steeply declining function of temperature in this temperature range. We exemplify this by indicating the factors by which the incorrect high temperatures used in {\protect\cite{Bulbul:2014sua}}  under-predict the K XVIII line flux compared to the predictions from the Ca XX to Ca XIX ratio.
}
\label{fig:caratio}
\end{figure*}

Point (2) in \citet{Bulbul:2014ala} intends to prove that there are no inconsistencies in the multi-temperature models adopted in \cite{Bulbul:2014sua} and that, in fact, we claimed such inconsistencies existed because of our use of WebGUIDE versus AtomDB. Here we clearly demonstrate that, as we originally argued, the temperature models employed in \cite{Bulbul:2014sua} are unrealistically biased towards large temperatures, and that they give inconsistent predictions for some key line ratios.

One of the most stringent tests of the self-consistency of a given temperature model is the use of line ratios corresponding to different transitions of the {\em same element}. Such a test obviously factors out the potentially large systematic uncertainty associated with relative elemental abundances. Unfortunately, \citet{Bulbul:2014ala} only quotes line fluxes for S XVI, Ca XIX and Ca XX (see their Table 2), so the only ratio we can use to test the self-consistency of the temperature models adopted in that study is Ca XX to Ca XIX. Luckily, such a ratio is rather temperature-sensitive, given the largely different peak temperatures of the Ca XIX and Ca XX lines.

In figure \ref{fig:caratio} we study the Ca XX to Ca XIX ratio for the Full Sample, MOS stack of clusters in \cite{Bulbul:2014sua}. The red line shows the predicted line ratio as a function of temperature (calculated using AtomDB v2.0.2), while the orange thick line shows the line ratio quoted in \cite{Bulbul:2014sua}. It is important to reiterate that the Ca XX to Ca XIX ratio is obviously independent of relative elemental abundances, and, at those temperatures, is a very direct probe of the plasma temperature. 

The Ca XX/Ca XIX ratio clearly points to plasma temperatures below 4 keV, while the set of temperatures in the multi-temperature model employed in \cite{Bulbul:2014sua} includes four components at 5.9, 6.1, 7.3 and 10.9 keV. Such temperatures are all grossly inconsistent with the one robust tracer of plasma temperature at hand. It should also be noted that for {\em all} samples and instruments in \cite{Bulbul:2014sua} the Ca ratio predicts temperatures between 3.1 and 4.2 keV, as we indicate with a red horizontal arrow in figure \ref{fig:webguide}.

It is important to appreciate that the impact of an incorrect temperature model biased towards large temperatures is even more dramatic when it comes to predicting the K XVIII flux: the latter is in fact a steeply declining function of temperature at the temperature of interest here. For example, as we point out in the figure, utilizing the Ca XX line flux to predict the K XVIII flux leads to {\em under-predictions ranging from factors of 4 to 13} when employing the temperature indicated by the Ca XX to Ca XIX ratio versus those used in \cite{Bulbul:2014sua}. Such large under-prediction factors go well-beyond the supposedly generous factor of 3 employed in the \cite{Bulbul:2014sua} fits, that would make their predictions ``maximally conservative''. 

Figure \ref{fig:caratio} also clearly illustrates  that the factor of 3 employed by \cite{Bulbul:2014sua} to account for systematic uncertainties from relative elemental abundances is not large enough to account for the incorrect and inconsistent temperatures employed in their prediction for the K XVIII, leading to under-estimates, for the case under consideration, of factors between more than 4 and more than 10.

Even in cases where a relatively low temperature component is included in the multi-temperature models, this component is typically sub-dominant in normalization compared to the high-temperature components. The systematic under-prediction of the K XVIII is thus present in those cases as well, casting serious concerns about the validity of the upper limits to the K XVIII emissivity employed in the fits of \cite{Bulbul:2014sua}.

Other aspects of the \cite{Bulbul:2014sua} temperature models are highly concerning. The sets of temperatures inferred for the {\em same} object with MOS and PN are markedly different: consider for example the ``Coma+Centaurus+Ophiuchus'' set: the temperatures inferred are:
$${\rm MOS:}\ 3.9,\ 6.8,\ 7.4,\ 10.7$$
$${\rm PN:}\ 2.5,\ 6.5,\ 7.2,\ 15.4$$
The resulting predictions for the K XVIII lines differ by a factor 2.6 (3.47 keV) and 2 (3.51 keV); the predictions for the K XIX 3.71 keV line differ by a factor of 4.1, and, finally, the prediction for the Ar XVII 3.62 keV line differ by a factor 9. Again, this applies to the {\em same sample of objects}, and goes to show how the multi-temperature models employed in \cite{Bulbul:2014sua} have large intrinsic systematics which have been not been accounted for in their analysis. Similar conclusions can be drawn for all samples utilized in \cite{Bulbul:2014sua}.

We reiterate that the key tenet of our original criticism to \cite{Bulbul:2014sua} regards the algorithm used to predict the range for the expected K XVIII flux, and the fact that such algorithm is  not conservative enough to allow for strong claims about the nature of the 3.5 keV line. As we have demonstrated here, and as also recently shown in \cite{Urban:2014yda}, such tenet holds and our concerns are valid.

In addition to the above concern about whether the allowed range of K XVIII fluxes employed by \citet{Bulbul:2014sua} is conservative enough, we would also like to point out that this is simply an {\it allowed} range, and in practice, the K XVIII fluxes do not typically saturate at the upper limits (A.~Foster, private communication).  It is therefore possible that some of the K XVIII flux is being associated, instead, with the purported ``excess" line.  While the ``excess" line energy found by \citet{Bulbul:2014sua} (3.55-3.57 keV) is somewhat offset from the K XVIII lines, given the instrumental energy resolution, the normalizations of these lines can easily be correlated.  The same is true of the Ar VXII line at 3.62 keV.  An important test which has not been done by \citet{Bulbul:2014sua} is to look at the covariance of the ``excess" line flux with the fluxes of nearby atomic lines.  A more conservative approach would be to enforce the neighboring atomic lines to be at their upper limits.  While the atomic lines are not necessarily this bright, this approach would give a better indication of whether an excess is truly present.

Finally, we welcome the additional search for the 2.96 keV Cl XVII Ly$\alpha$ line, an obvious cross-check on whether the 3.51 Cl XVII Ly$\beta$ line could or not contribute to the 3.5 keV signal that we could not have performed given that \cite{Bulbul:2014sua} did not quote limits on the 2.96 keV line (see also however, \cite{Urban:2014yda} who do claim the detection of Cl in the Perseus cluster). In no cases, however, did we predict a dominant Cl XVII contribution to the 3.5 keV line compared to the two K XVIII lines: assuming solar abundances, the relative contribution of Cl XVII compared to K XVIII is always on the order of a few percent at most, especially at low plasma temperatures. This new result is therefore also inconsequential to our findings and conclusions. 

\section{Concluding remarks}
In this short note we have demonstrated that all three key points we originally made in \cite{bananas} are valid and that the criticisms put forth in \cite{Boyarsky:2010dr} and \cite{Bulbul:2014ala} are either incorrect or inconsequential to our results.

We demonstrated that the occurrence of a 3.5 keV line in the M31 XMM data is due to poor background modeling when utilizing an excessively broad energy window; the lines purportedly detected in \cite{Boyarsky:2014paa} are very likely spurious, and are physically inconsistent among each other. We showed, for example, that when utilizing the broad 2-8 keV energy range, an unphysical absorption feature at 2.9 keV improves the fit by a much larger factor ($\Delta 
\chi^2 = 52$) than adding a line at 3.5 keV ($\Delta 
\chi^2 = 12.5$), while also reducing the significance of any excess line near 3.5 keV to about $2 \sigma$. We thus confirmed that no statistically significant 3.5 keV line exists in the X-ray spectrum of M31, and explained the origin of the features reported in \cite{Boyarsky:2014paa}.  Fitting an energy window vastly larger than the instrumental energy resolution (3-7 keV), a 3.5 keV line feature is not needed or significant.

We demonstrated that the multi-temperature models employed in \cite{Bulbul:2014sua} are inconsistent with line ratios independent of relative elemental abundances, and that they are  systematically biased towards often unrealistically large temperatures. As a result, the predictions for the K XVIII line flux are systematically underestimated, in a way that casts significant concerns on the validity of the results  presented in \cite{Bulbul:2014sua}.

We suggested to Bulbul et al an important cross check on the cross-correlation between the nearby K XVIII and 3.62 keV Ar XVII line fluxes and the putative additional 3.5 keV line. Should evidence for significant cross-correlation exist, and in light of additional proof of under-estimated upper limits to the expected line flux, the case for a K XVIII line as the origin of the 3.5 keV line would be further strengthened.

In closing, we note that our recent study on the morphology of the 3.5 keV emission from the Galactic center and from the Perseus cluster of galaxies \citep{carlson} conclusively demonstrates, using an entirely orthogonal method to spectral analyses, that a dark matter decay or axion-like particle conversion origin for the 3.5 keV signal is robustly excluded.

\section*{Acknowledgments}
%
\noindent We thank E.~Storm, A. ~Boyarsky, and A.~Foster for helpful discussions.  SP is partly supported by the US Department of Energy, Contract DE-SC0010107-001. 
%

\bibliography{galcenter}

\end{document}